# ENBB Processor: Towards the Exascale Numerical Brain Box[1]

[Position Paper]

Elisardo Antelo

Dept. Electrónica e Computación

University of Santiago de Compostela

SPAIN

elisardo.antelo@usc.es

November 22, 2012.

*Abstract*

ExaScale systems will be a key driver for simulations that are essential for advance of science and economic growth. Current technology trends indicate that there might be a big energy wall by the end of the decade. Different reports call for strong changes at all levels for ExaScale computer systems. This academic position paper addresses this problem in the context of the microprocessor, the key element for performing numerical work. We aim to present a new concept of microprocessor for floating-point computations useful for being a basic building block of ExaScale systems and beyond. The proposed microprocessor architecture has a front-end for programming interface based on the concept of event-driven simulation. The user program is executed as an event-driven simulation using a hardware/software co-designed simulator. This is the flexible part of the system. The back-end exploits the concept of uniform topology as in a brain: a massive packet switched interconnection network with flit credit-based flow control with virtual channels that incorporates seamlessly communication, arithmetic and storage. Floating-point computations are incorporated as on-line arithmetic operators in the output ports of the switches as virtual arithmetic output channels, and storage as virtual input channels. The front-end carries out the event-driven "simulation" of the user program, and uses the arithmetic network for the hard floating-point work by means of virtual dataflows. We would expect to reduce significantly the needs of main memory due to the execution model proposed, where variables are just virtual interconnections in the network or signals stored in the virtual channels. Moreover, we have the hypothesis that the problem size assigned to a microprocessor should allow maximum concurrency and it should not be oversized. This may lead to systems composed of microprocessors with main memory incorporated in 3D chips. We identified several challenges that a research to develop this microprocessor should address, and several hypothesis that should be demonstrated by means of scientific evidence.

*1-Introduction*

High-end general purpose microprocessors have to evolve to cover two specialized paths: intelligent computing and numerical computing. Intelligent computing is not related to raw floating-point performance, but to new algorithms to have real intelligent machines. Strong floating-point computation performance will continue to be essential for simulations in science and engineering. In fact, these two paths of specialization

---

[1] This work was developed while the author was with the Politecnico di Torino (Torino, Italy), during the summer of 2012, supported by a grant from the Goverment of Spain. I thank all the support and feedback provided by Prof. Paolo Montuschi while I was with the Politecnico di Torino.

have already emerged with GPU-like processors for high-performance computing, and initiatives such us IBM Smarter Computing (an instance is the Watson computer)

Today's science and the innovation driven industry depend on a continuing increase in computation performance at affordable cost (i.e. energy costs). Simulation is the key tool for research and development in many areas: computational biology, material sciences, nuclear engineering, new energy systems, combustion systems, weather prediction and climate modeling, physics, aerospace technology, security, etc. This trend is expected to be even stronger during this decade to keep economies competitive.

In recent years there has been several efforts to trace a path to ExaScale supercomputers [1-4] (10 floating-point operations per second) constrained to 20 MW of power consumption by the end of the decade from the current 10-20 PFs ($10^{18 15}$ floating-point operations per second) at about 8 MW of power. The problem is not the ExaScale level of computation, but the power constraint at this level (energy efficiency).

Several studies [4] remark the economic, social and even geostrategic importance of achieving this milestone, and the extraordinary research effort required at different levels. Among the fundamental problems, energy consumption is the key issue. Microprocessors, main memory, interconnection network and massive storage should be much more energy efficient to achieve the 20 MW goal. This energy efficiency can be expressed in simple terms: about 50 times more computational performance with only 2.5 times more power consumption. An evolutionary path of the current technology seems not to be an option, and deep changes at all levels will be required.

In this context, for this position paper I concentrate on the microprocessor, the true computation engine. Specifically, the goal of this paper is to present an initial proposal for the architecture of the "***ExaScale Numerical Brain Box***" or **ENBB** for short. "ExaScale" because is intended to be part of future ExaScale systems and beyond. "Numerical" because we intend a design for numerical floating-point computations essential for the high performance simulations that now science and engineering needs. "Brain" in the sense of energy-efficient processor. "Box" because integration in the three dimensions is the path to follow as it is on the brain. This architecture, based on a new concept of data flow network computing on a chip, is far from the path industry and academia are following, but I will try to provide enough arguments to justify this drastic change. Moreover, for the post ExaScale era, this proposal may open the path to new computer architectures for hard floating-point computation with post-CMOS technologies.

*2- State of the Art*

CMOS technology has enabled an exponential performance scaling but at the cost of increasingly higher energy and power. For the next generations of silicon processes it is expected that the number of transistors doubles each generation [5] (Moore's Law continues), while according to not optimistic predictions [5-7], energy per switch will scale down only by about 25-35% per generation due to the difficulties in scaling the power supply voltage and the active capacitance. Therefore, most experts predict that an ever increased fraction of silicon in future microprocessors will be "dark" [6], that is, it will not be possible to fully use all the transistors available on the chip due to power and energy constrains.

There are two basic commodity microprocessor platforms used today for high performance computing: general purpose high end processors (GPPs) and graphics processors (GPUs). GPPs have typically 8-16 cores of significant complexity (although there are variations depending on the vendor), with a frequency of 2-3GHz, large on-chip last level caches (up to 24-32 Mbytes), and high memory bandwidth (about 50 Gbytes/s for local memory and about 100 Gbytes/s for remote higher latency memory). Each core supports 2-4 logical threads in hardware. The thermal design power for this kind of microprocessors reach 130-180 W with a peak performance of 60-100 GFs ($10^9$ Floating point operations/sec). GPUs have a huge number of small cores (2000+) supporting a huge number of threads, with frequencies around 0.7-1.0 GHz, internal storage for streams (local memories), and much higher memory bandwidth than general purpose processors (about 200 Gbytes/s for local memory). The thermal design power for GPUs may reach 200-300 W, with a peak performance approaching 1 TFs ($10^{12}$ Floating point operations/sec).

These two kinds of platforms favor applications of different characteristics: GPPs favor applications with high levels of data locality and a reasonable amount of parallelism, being the latency of threads an important factor; GPUs favor applications with minor levels of data locality, a huge level of data parallelism, being the throughput of threads an important factor. In general GPPs are easier to program than GPUs, especially in the case of irregular applications. Both platforms suffer from the memory wall problem: GPUs try to solve it by using massive multithreading to hide memory latency but using a very high memory bandwidth; GPPs rely on large caches and modest levels of core multithreading. Both platforms take advantage of reduced memory latency by integrating the memory controllers on chip. Both platforms try to boost performance for applications that do not match well the processor architecture. Thus, GPUs incorporate small caches and GPPs incorporate SIMD datapaths and instructions for software prefetching.

In current high-performance systems the trend is to have heterogeneous computing nodes that combine both types of platforms, some GPPs and some GPUs as coprocessors, to have both flexibility and top floating-point capabilities [2]. Future plans for microprocessor vendors for high performance computing show the convergence of this kind of node in a single chip heterogeneous microprocessor with some general purpose huge cores and many small GPU type cores [7-9]. This trend is driven by the exponential growth in the scale of integration of transistors. Another trend is to manage the frequency/voltage of the cores individually to adapt the power consumption to the needs of the computation. Cores can also be power gated.

During the last years there has been a trend in the design of highly energy-efficient top supercomputers. Specifically, the BlueGene series of supercomputers and the K supercomputer are clear examples. In these cases microprocessor vendors adapt a line of general purpose microprocessors to the specific power-performance needs of supercomputer applications.

On the software side, there are two traditional paradigms for concurrency programming: shared memory (with APIs such as OpenMP or Cilk) and message passing (with APIs such as MPI). The shared memory paradigm implies that the concurrent threads of a program communicate by accessing shared memory positions. The programmer has to take care of possible data races and the necessary synchronization. The need for efficient private caches in a multilevel cache hierarchy may lead to coherency problems. Shared memory systems usually support hardware coherence to solve this problem in a way almost transparent to the programmer (but with side effects such as false sharing). The programmer should also take care of the consistency model (order of reads and writes of different threads) supported by the programming tools and hardware. For the message passing paradigm the concurrent processes communicate through explicit messages that the programmer includes in the code (blocking or not blocking point to point or collective).

These paradigms were used traditionally to program multiprocessors, however the model is extensible to program multicore microprocessors. Both paradigms can be used together when programming a system with shared memory nodes (several sockets with multicore processors) connected by a non-coherent high speed network such us Infiniband (some languages support both paradigms directly). Some industrial prototypes indicate that the trend for multicore processors could be an hybrid between message passing and shared memory, since full true shared memory might result very costly in terms of energy and power.

To program GPUs (with APIs such as CUDA or OpenCL) the usual model defines kernels to be executed in a parametric form with respect to the data (some sort or simple kernel multiple data). During execution, each instance of the kernel leads to a hardware thread in the GPU. Most of the data movement (usually streams) from/to memory is responsibility of the programmer.

*3- Main Threats to this Model for the ExaScale Target*

There are many threats to this model, most of them related to power and energy efficiency. At the time of writing this paper, the most power efficient microprocessors for supercomputing applications achieve about 2-4 GFs/W, or equivalently, 250-500 picojoules per floating-point operation (pj/Flop). Current GPU-like processors may achieve theoretically about 250 pj/Flop, but the difficulties to achieve a high utilization in general may lead to lower efficiencies.

Efficiencies at the system level are reduced because of the energy needed for memory, massive storage systems, communication between microprocessors and cooling. For instance, the supercomputer BlueGene/Q achieves about 20 PFs of peak performance (16.3 PFs sustained performance for the benchmark used in the Top500 list) with a total power of about 8 MW. This leads to about 2 GFs/W, or 500 pj/Flop. The BlueGene/Q microprocessor [10] achieves a raw efficiency of 3.7 GFs/W or 270 pj/Flop. This means that about half of the energy is spent in moving data between microprocessors (network), storage and communication with main memory, massive storage and cooling. A typical double precision FPU (Floating-Point Unit, usually a fused multiply-add unit) may consume 15-30 pj/Flop in the same technology. This means that inside the microprocessor, only about 5-10% of the energy is devoted to actually performing the floating-point operations (the "golden" work). This is reduced to 3-6% for the whole system.

An ExaScale supercomputer with a target of 20 MW (this was established by the US Government in some of their supercomputer programs and it is now widely adopted as a target for ExaScale), will need an efficiency of 50 GFs/W or 20 pj/Flop at the system level. Very optimistic scaling scenarios (ITRS scaling roadmap, with aggressive scaling of power supply and active capacitance per unit area) state that energy per operation may scale by a factor x0.5 per CMOS process generation. More realistic predictions [7] reduce this scaling factor to x0.7 per generation. Assuming an ExaScale system in four generations, leads to a x16 energy reduction per operation in the optimistic scenario and about x4 in the more realistic scenario. Therefore we could expect 1-2 pj/Flop for a scaled FPU being very optimist, and more realistically 4-8 pj/Flop. This leads to 5-10% of total energy at the system level for pure arithmetic work in the optimistic case, and 20-40% for the more realistic scenario (**in this case the arithmetic work may represent 35-70% of the total energy at the microprocessor level**).

The optimistic scenario leaves in hands of technology scaling the path to ExaScale, with evolutionary optimizations in different parts of the system to achieve the savings of energy that would be required to host the shift from 3-6% to 5-10% of relative contribution to total energy of the arithmetic work. There are some interesting results already for this scenario with the design of microprocessors with an aggressive scaling of the power supply to reach the near threshold voltage zone [11-12]. It remains to be seen if the near threshold approach is a viable solution for mass production and not just for prototypes. This low level of voltage raises many issues related to reliability. Moreover, the near threshold approach makes the cores very slow (although very energy efficient), so that for a required performance the parallelism needed (number of cores) is much higher, putting more pressure on parallel programming (see below) and in area. Another issue is that for a given power envelope, the electric current should raise by more than a factor of two, leading to the need of much more pins of the chip devoted to power and ground (less pins for data bandwidth).

The more realistic scenario predicts a shift from 3-6% to 20-40% of relative energy contribution for arithmetic work at the system level (**a shift from 5-10% to 35-70% at the microprocessor level**). This would imply deep changes of design to reduce significantly relative energy contributions of non-arithmetic work.

Interconnect between computation nodes are much more power and bandwidth efficient with optical interconnect, so that this technology is already being used. However, there are many industrial and academic projects that promise optical interconnects even at a short distance level, or even inside the microprocessor chip. Silicon photonics intends to provide the integration of optical interconnects in silicon CMOS processes. This may solve the problems related to the energy of interconnects out of the chip, and the bandwidth wall.

The memory system is also a concern in terms of energy, power and storage density. Current memory systems are designed for low cost storage, and they are not suitable for ExaScale computing. However, also in this area there are very promising results from several industrial and academic projects. Among some current developments, the most important are: i) the transition to 3D integration of the memory chips (memory cubes) that will reduce the footprint of the memory system, with shorter distances to the microprocessor and reducing power due to a higher level of integration; ii) the possible transition to a *memristor* like storage devices (i.e. Phase Change Memory) that may solve the serious problems of scalability of capacitor based storage cells.

Regarding the microprocessor, there are several ongoing industrial and academic research efforts that try to reduce the relative energy overhead of non-arithmetic vs. arithmetic components (see [7] [13] and [14] among others). The main efforts are on increasing the levels of locality in the computations, to avoid unnecessary data movements: moving four 64-bit operands of data a length of 1mm may take about the same energy as performing a double precision floating-point fused multiply-add operation involving those four operands. In order to improve locality and to reduce also the use of the external links, there are proposals for having configurable internal memories that may be used as caches or software managed memories, or even register files. The design of power-efficient cores, the balance of complex vs. simple cores (heterogeneous system), the chaining of operations inside the cores, the design of power-efficient on-chip interconnection networks, the design of efficient hardware support for thread synchronization, are under deep research. At last, the objective of these research efforts is to devote more silicon to floating-point operations and less to control structures related to the decoding and execution of instructions and to internal networks for data movement. Technology advances such us board level optical interconnect and 3D integration will also make a strong contribution for energy efficiency and performance [15].

A second major threat is the increasing need for reliability support and resilience in general. An Exascale supercomputer will be of extremely high complexity in terms of components of all types, and many of those components will not improve their reliability unless specific actions are taken. Therefore the failure rate will increase drastically, so that further strong research is needed for not having mean times to interrupt of minutes. The main reason for lower reliability of components is the increase in manufacturing variance, lower voltages, and smaller transistors and interconnections. Current approaches that rely on fault detection and checkpoint/restart need to evolve to be much more efficient. Therefore, regarding the microprocessor, increasing levels of fault tolerance (evolving from current practices of fault detection: ECC memory protection, parity protection of register files, residue checking in floating-point units) will be necessary.

A third major threat is the programming of highly parallel systems. The main concerns are the limited speedup due to Amdahl's law, the low productivity of parallel programming and the difficulties in verification of parallel programs due to non-determinism of the execution model. There are several industrial and academic efforts that try to mitigate these problems. However for scientific and engineering high performance simulations this is a second order concern compared to the energy and reliability issues.

*4- The ExaScale Numerical Brain Box*

Our specific goals and priorities are the following: i) energy efficiency computations are the top priority; ii) provide a programming model tightly coupled with the proposed architecture to allow energy efficiency and effective parallel programming is a secondary priority; iii) contribute to improve the reliability of the microprocessor is a goal of moderate priority. This order of priorities is justified by the fact that energy efficiency at the microprocessor level is unavoidable to reach the ExaScale goals. Technology advances will help, but architecture innovation is an orthogonal effort to technology, and could provide further reductions in energy and power consumption, that is always a positive competitive factor in an era of expensive energy. The other issues, such as parallel programming and reliability, although very important, may find complementing solutions at levels higher than the microprocessor hardware.

We already mentioned the human brain as part of the name of our processor. The brain represents the path to follow in terms of energy efficiency. Although very different to a numerical processor, we would like to incorporate to our architecture any general "design philosophy" of the brain useful for energy-efficiency.

**4.1 A New Path for Data is Necessary**

After reviewing many published proposals (academic and industrial) for future microprocessors for the ExaScale era, I realized that none put in question the model of having data parallel execution units, separate and hierarchical on-chip storage (register files, caches, local memories and queues), and data parallel movement of data (on chip: network on chip, and off-chip: DRAM interface). These layout elements "were", "are" and it seems that "will" be an essential and unquestionable part of microprocessors. For this position paper I put in question this standard practice and look for aggressive and risky alternatives.

Current parallel floating-point units for floating-point fused multiply-add are very complex datapaths: for double precision require internal datapath elements and interconnections up to +100 bits wide. Moreover, the pipelining of these complex units requires a significant number of flip-flops, since these circuits are very wide. These units are designed to reduce latency. The need to support different formats in a SIMD form lead to even more complex structures. These data parallel units have their own design space in terms of latency, energy, power and area, but the options are severely affected by the underlying data parallel architecture.

Register files are the fastest storage elements in the microprocessor. These units have several read and write ports to support the interface with the execution units and the higher levels of storage (cache or local memories). There is a continuous need to increase the number of registers to support more parallel execution leading to very complex hardware structures. Regarding on chip interconnection, close to the execution units there are wide buses to communicate with the register files, and there are also wide buses for feedback loop on those units. Moreover, in the path to integrate more cores in a chip, networks on chip are necessary (rings, mesh,..). Again, these networks support wide data parallel (up to 512 bits) movements that make interconnect layout and routers very complex. Finally, the integrated memory controllers interface with the DRAM chips by means of wide data parallel channels (in the standard DDR-x form, 64 bits of data but with a total of 240 signal pins). Some advanced interfaces use narrow serial channels, but inside the chip, the data is converted to data parallel form. Therefore, we conclude that data parallel microprocessor architectures are driven by the need of low latency computation and communication. These needs are in part due to the instruction set architecture and the programming model of current microprocessors.

An empirical law for VLSI circuits states that A x Perf$^2$=constant (A: area of the circuit, Perf: performance of the circuit). This means that a circuit with four times less performance than a reference circuit, may occupy sixteen times less area. The loss of performance may be compensated by using four instances of the lower area circuit with still four times less area. This is exactly the same argument for using many small cores instead of few big cores (Pollack's rule). Of course, this strategy has practical limits in terms of the need of higher parallelism and the overhead of managing more units. On the other hand, since energy is the product of the active capacitance times the square of the voltage, a first order approximation leads to identify area and active capacitance, and therefore to identify area and energy. Thus the relevant metric of energy per floating-point operation can be expressed as the reciprocal of the metric (Flop/cycle)/mm$^2$ (floating-point operations per cycle and per unit of silicon area) and probably, (Flop/cycle)/mm$^3$, taking into account future mainstream use of 3D integration [15]. Therefore, energy per floating-point operation is minimized by maximizing (Flop/cycle)/mm$^3$.

These ideas suggest that for energy efficiency, it is better to have more and simpler floating-point units. Moreover, as discussed earlier, communication is very costly (in comparison to floating-point operations), and storage is also very costly, and in many cases redundant (for instance, in the case of inclusive caches). The area/volume and energy devoted to separate communication and storage is not used to perform floating-point operations, leading to a decrease of the metric (Flop/cycle)/mm$^3$. All these arguments will be even stronger with increasing levels of precision (quad precision might be necessary for Exascale systems).

A further argument against separate computation, communication and storage is related to the topology of the neocortex in the brain (our reference for energy efficiency). The standard topology of a microprocessor is heterogeneous, with different parts of the chip devoted to computation, communication and storage, while the whole neocortex (a 3D structure of about 1000 cm$^2$ of surface and 2 mm thick) is roughly uniform, according to some evidence, basically an interconnection network of neurons organized in layers and columns. This might not be considered a strong argument, since the neocortex implements very different "computations" than a microprocessor, i.e., the uniform topology could be good for the workload of the brain, but not for floating-point computations. Moreover, the basic elementary components for communication and processing are very different. However, it is evident that the separate topology in microprocessors is not providing an energy efficient solution and this is one of the reasons why the model is in crisis, and alternatives should be found, and the neocortex looks a very good reference.

Based on all of these arguments, we have in mind an architecture with a uniform topology for processing the data. Specifically, we want to design a throughput oriented **architecture that integrates seamlessly arithmetic, communication and on-chip storage**. For energy efficiency we want to make a massively (highly parallel) use of serial computation and communication, leading to an energy efficient computational network. This concept resembles what is done in some supercomputers (for instance the K and BlueGene/Q supercomputers) for reduction operations that are performed on the external network components.

We envision a massive network of nodes, with narrow serial interconnects between them in a 3D mesh topology (or other topology if it is more efficient) spawned in the three dimensions of the chip. Floating-point data flow into the nodes and results flow out as a combination of some of the inputs by means of an arithmetic operation. For a perfect fit of the arithmetic computation and the network, we will use on-line most significant digit first arithmetic units (on-line arithmetic for short).

**On-line arithmetic [16]**: this is a kind of serial implementation of the arithmetic operators that overlaps computation and communication. The digits of the operands input the unit in serial form (digit-by-digit, one digit per cycle) starting from the most significant. After a small number of cycles (the on-line delay), the digits of the result are produced in the output in serial form. The output digits are ready to input other on-line units. This is possible because, for on-line arithmetic, the representation of operands is redundant to avoid the need of carry propagations. Therefore, by chaining several operations, it is possible to have a powerful computation engine composed of low-cost serial units. The key point is the overlap between computation and communication: the delay for the first digit in a chain of operations is only the addition of the small on-line delay cycles of each of the component operations.

This kind of arithmetic is not used in microprocessors today. In fact only a few publications report the use of on-line arithmetic for implementing application specific circuits for applications such as communications or matrix algebra. The development of on-line arithmetic dates back from the mid-seventies, with several papers published at that time. Since then, only a few authors made some research in this area, leading to a sparse number of contributions. The lack of interest is motivated by the low practical industrial applications, in part due to the prevalence of low latency oriented data parallel arithmetic, so that it remained as a niche research topic in computer arithmetic. Although fixed-point algorithms for basic operations are well known, the corresponding floating-point implementations still present some challenges. Recently there have been proposals for floating-point on-line units for addition and multiplication, that solved some of these challenges [17]. However, further research is necessary, especially in relation with the specifications of the IEEE 754-2008 for Floating-point arithmetic and the design of fused addition multiplication units.

For high utilization and flexibility, we intend to use a **packed switched network**, using some sort of flit credit-based flow control with virtual channels. The use of a network of on-line arithmetic elements is not new. In [18] authors propose a reconfigurable hypercube network of on-line arithmetic units for the evaluation of complex arithmetic expressions. They concentrate on the mappings of different computation tree graphs into the hypercube network. In this proposal we go a step further by using a packed switched network as a computational engine, integrating seamlessly chip-level arithmetic, communication and storage. **This is a key element of our architecture**. The virtual channels (storage in each switch) should provide the necessary support for the control flow policy (communication), and also would be used as internal storage of data for the microprocessor. A problem related to on-line arithmetic is that due to the redundant representation of operands, the storage needed is higher (more bits to represent a data format). Therefore the specific representation selected might be critical for the resultant energy consumption. On the other hand, the storage of virtual channels could be integrated with low cost and low power storage elements since our architecture is throughput oriented, less dependent on latency issues as in current architectures.

In this system operands may move on the network in a serial form using virtual channels in each hop, going to a destination node where are combined with other operands to create a new stream of digits (a new operand) that outputs digit by digit with destination to another node for further combinations. Therefore, the output ports of switches may forward one of the inputs or combine some of them by means of an on-line arithmetic operator (output arithmetic virtual channels). Thus, the network creates new operands, and may destroy old operands if they are no longer needed for further computation. In summary, operands may be created, destroyed on the network, and along its life, an operand will "visit" nodes in serial form to generate new operands or to provide control information.

The operands will be embedded in packets with additional information for flow control, routing, etc., and will be decomposed in the basic units of information for flow control (flits), in such a way that during the trip to destination, the operand is distributed, with flits stored in different virtual channels of different switches along the path. This architecture for the datapath has the potential advantage of simplifying the methods needed to assure reliability, due to the uniform topology. Moreover, the reliability and the performance of the microprocessor will be tied in part to the quality of service of the on-chip network.

Our hypothesis is that a datapath based on these ideas may lead to a very energy-efficient design. Deep research will be needed to provide scientific facts to demonstrate this hypothesis and to overcome the challenges.

Specifically, an incomplete list of challenges is the following:

a) The representation of the operands for on-line arithmetic is redundant. This leads to several problems, such us increasing the storage size per operand and difficulties in fully implementing the requirements of the IEEE 754 2008 Floating-Point Standard. Future technologies may help, such as *memristor* type memories that allow multi-bit storage cells. Future post-CMOS technologies may also help, since most of them support multilevel logic naturally.

b) Most significant digit arithmetic is not appropriate (long on-line delay) for operations such as sign detection, absolute value or the calculation of remainders of division operations.

c) As in conventional networks, we must take care of deadlocks. The on-line arithmetic operators that create new operands in the network might be an added problem.

d) The floating-point computation may require some support of integer operations (not related to control or computation of addresses, since this will be handled in a separate module as we describe below). This support should be embedded in the configurable floating-point on-line operators.

e) A complete design space should be explored: specific topology of the network, radix of the switches (number of input/output ports), number of virtual channels per physical channel and configurability of the on-line arithmetic operator in the output ports, even providing several virtual computation channels at the output to improve total network throughput..

f) The network will host movement, storage, creation and elimination of operands. Those actions are highly correlated with the actions described by the input program (control of the computation). The network will need a sort of instruction level architecture (control) to interface with higher level structures (to be described below) to implement the actions required by the input program. For instance, control packets could flow to reserve virtual channels for implementing a whole kernel.

g) Study in depth the reliability issues of this architecture. Moreover, interval arithmetic is of much interest for numerical simulation since allows checking the validity of results and may allow using more aggressive parallel algorithms. Could our datapath architecture provide hardware support for interval arithmetic in an efficient way?

h) On-line arithmetic allows taking advantage of using only the "correct" digits of an operand (this is not the case in data parallel units). This may allow important optimizations when the programmer is aware of the number of digits that are suitable for each of the operands[2]. More research would be needed to determine how to support efficiently this scheme.

Figure 1 illustrates a high-level view of the switch for the arithmetic network. At the input we have virtual flit channels and at the output the arithmetic on-line channels. Some shadows flit buffers would be necessary to keep a copy of an operand while sending the operand to an output destination (for instance, for the case when a computation has a fanout greater than one, or when the operand is combined with other operands by means of an arithmetic operation to produce a new operand).

Figure 2 illustrates a computation in the network, specifically two operands A and B are combined by means of an arithmetic operation to produce operand C. The operands are distributed along the switches of the network and flow on a flit basis.

---

[2] This idea was raised by Florent de Dinechin, ENS Lyon during a talk about the ENBB microprocessor.

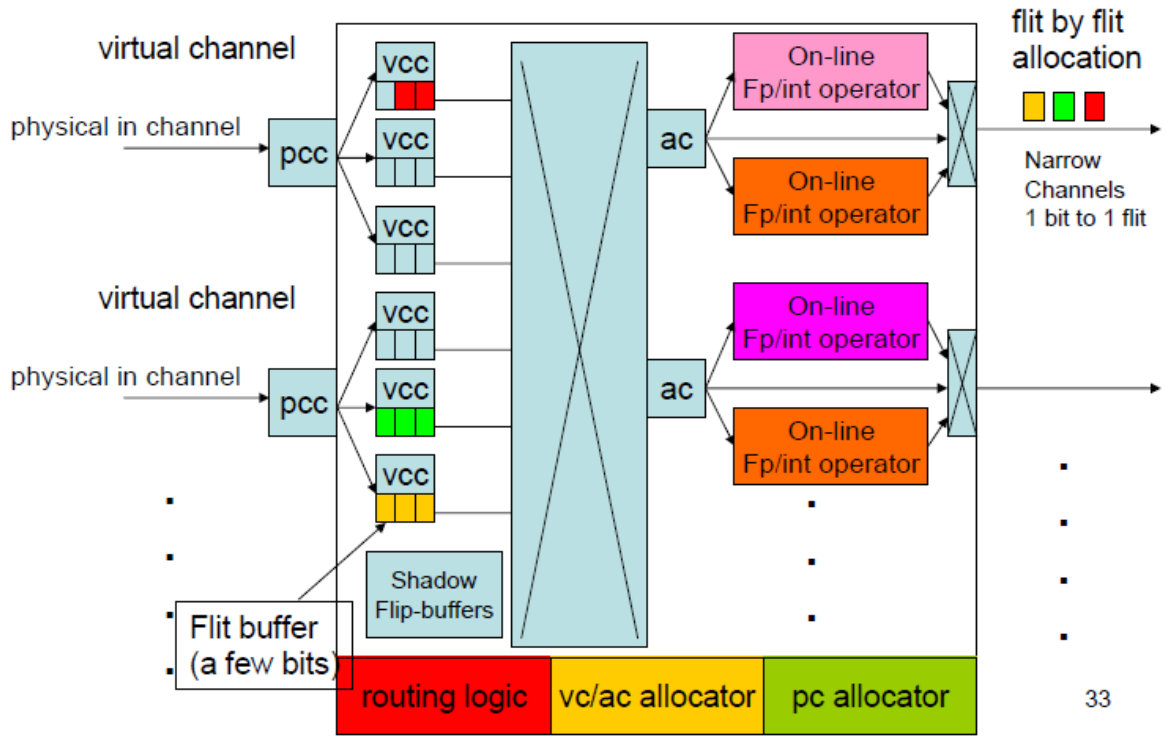

**Figure 1: Switch architecture of the Numerical Brain.**

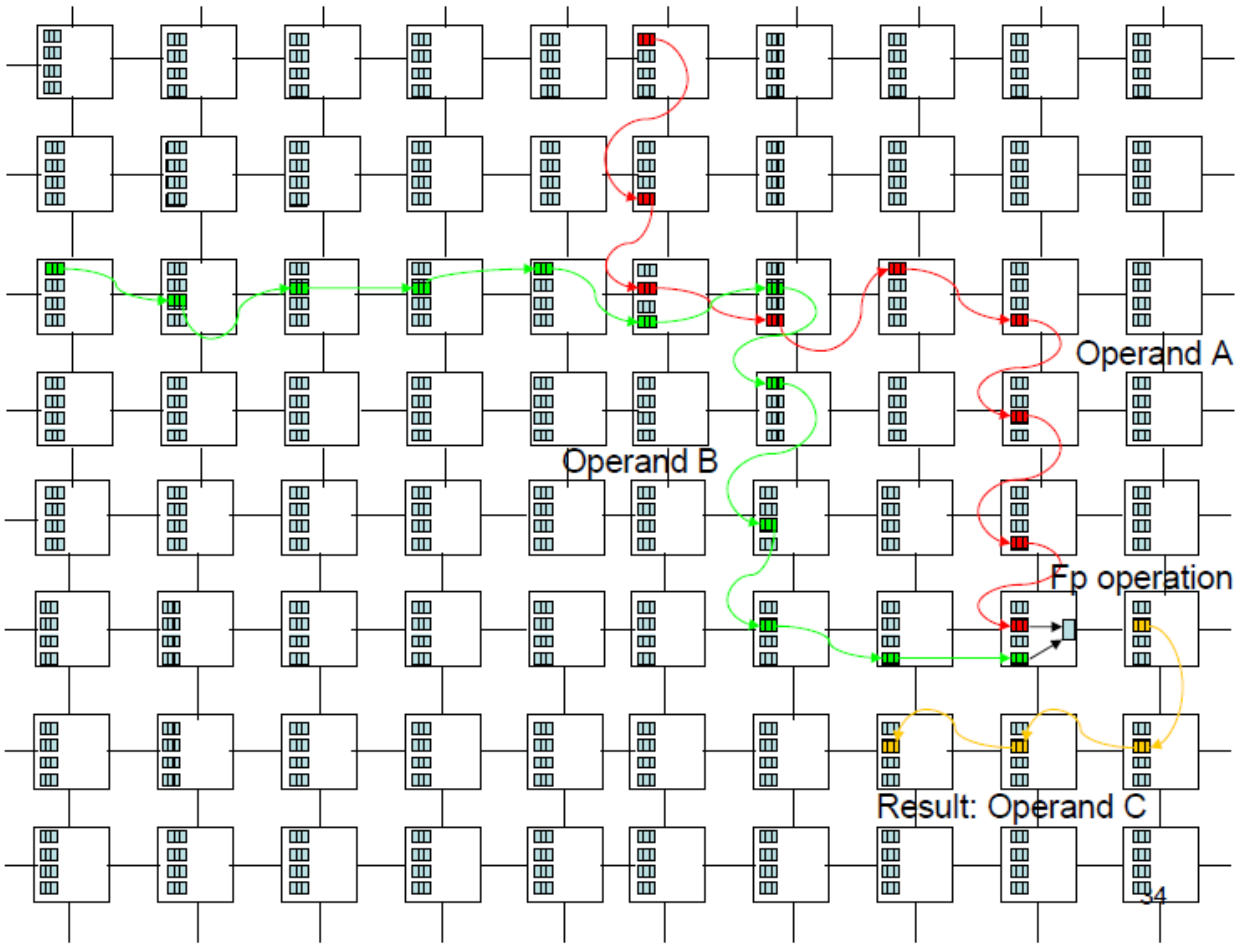

**Figure 2 : Instance computation performed in the Numerical Brain.**

*4.2 Less (Off-Chip Memory) Might be More (Efficient)*

We propose another interesting hypothesis that should be demonstrated. Our hypothesis states that having much less off-chip memory per computational node might be globally more efficient. Current practices scale linearly the off-chip DRAM memory capacity with microprocessor performance. Following the current trends and technology, there are several issues with the memory systems for ExaScale as we discussed before. Even with new technologies (memory cubes, *memristor* type memory cells and optical interconnect), it would be desirable to scale down the off-chip memory needs: less energy (less chips, less interconnect, less footprint of blades) and less impact on performance (memory latency and bandwidth).

I argue that there are three strong drivers for high requirements of off-chip memory: i) a significant part of all variables in programs have off-chip memory allocated during a significant part of the execution time, in part due to the programming paradigm used; ii) to support multiple processes in time sharing form; and iii) usually, in a parallel execution each of the compute nodes are assigned an oversized part of the problem, so that the computing engine is not able to concurrently perform the assigned work, needing to store more information. This might have sense for general purpose computing, however, we argue that this is not necessarily the best option for high performance computing. Our objective is to conceive a programming interface (we address this issue in the next subsection) that reduces the needs of external off-chip memory. Moreover, we intend to propose a programming interface to get the best performance when the size of the problem assigned to a microprocessor allows the use of maximum concurrency (not oversized). If the problem size for a compute node is too big, then performance would be degraded, since a more frequent access to a slow and low energy secondary memory would be needed. We want to make low memory requirements the common case. This would allow to remove most of the needs of energy for the off-chip memory infrastructure (memory chips and interconnect).

The maximum concurrency paradigm is inspired by the processing performed in the neocortex. According to some evidence, it seems that the neocortex processes the sensory data with maximum concurrency to detect and predict patterns (all the stimulus at a time), and memory is used only as part of the computation (storage of patterns) and not for serializing the processing of the arrived data at a given time.

In an optimistic scenario, all the main memory needed would be integrated in the same "box" with the different layers of the microprocessor, that is, some layers of the 3D chip would be devoted to high density memory [15]. The high level of integration of a compute node in a box would allow using more compute nodes under the same energy envelope, with each compute node having a problem size that allows maximum concurrency on chip. We hope that removing board level components (memory chips and fast off-chip interconnections) will allow using enough compute nodes to host the problem sizes needed. We have the hypothesis that this strategy is feasible and that it may produce very good results without limiting problem sizes. The challenge is to provide scientific demonstration of this hypothesis.

*4.3 A New Programming Interface for the Numerical Brain*

The datapath introduced in the previous sections requires a control front-end conceptually different from the current practices (a separate farm of superscalar front-end on cores). The project must provide the control interface and a first layer for programming. We restrict our research to the microprocessor level, although a conceptually similar scheme could be scaled to multiprocessors.

Our datapath fits naturally to what is known as the dataflow model (in contrast to the control flow model). Dataflow machines [19] had some popularity during the mid-seventies and eighties, even with some commercial instances. The dataflow model is based on the execution of operations driven by the availability of data and following the dataflow graph described by the input program. The dataflow model has the potential of achieving energy efficiency, since it tries to make an efficient mapping between the higher level programming language and the hardware. ASIC implementations might be considered an extreme example of dataflow machines, since the hardware is fully adapted to a specific "program" (no flexibility), being between one a two orders of magnitude more energy-efficient than standard microprocessors. Systolic arrays could be considered another instance of the dataflow model, although they are highly specialized for particular applications. The dataflow model is now receiving again certain attention due to the need of higher levels of energy efficiency.

From my point of view, the dataflow model has the following problems: i) it is too much focused on obtaining a direct dataflow graph from the high level programming language (or an ISA) finding the typical problems that the hardware synthesis community find when tries to synthesize hardware circuits using high-level design languages, ii) the underlying hardware is very close to the control flow architectures, and even now the dataflow model is mapped directly in conventional multicore systems with runtime support.

For specific problems, dataflow may lead to very good results. One example is the high performance computing engines based on dataflow provided by Maxeler, where a static dataflow is mapped to a configurable hardware (data parallel dataflow engines) to implement a simple kernel multiple data model. Static dataflow is also used for network processors. Similar schemes are used at the core level to chain several operations using a static circuit switching network [13]. The problem is that for more flexibility, dataflow should be dynamic, i.e., data-dependent data flows, that are much harder to support.

Our project aims to define a simple and flexible programming interface for parallelism that is suitable for our novel datapath architecture. We hope that the tight design of datapath, control and programming interface lead to higher energy efficiency.

We agree that the dataflow computation model is very attractive, but more flexibility and abstraction is needed. Our starting point is a class of languages that allow an event-driven paradigm. Although many alternatives exist, to focus the discussion we concentrate on languages used to model hardware, specifically we concentrate on VHDL-line languages. Note that we are only interested in the event-driven execution part of the language and not the high-level features that allow flexibility in programming, since VHDL-like languages are recognized as limited in high level features.

VHDL-like languages have been used for more than two decades to model and/or synthesize hardware circuits of very high complexity. Specifically we are interested in the simulation paradigm. VHDL-like simulators take the input program, a sequence of input test vectors and perform a discrete event data driven simulation. We want to keep simplicity and use well proven practical concepts, so although this kind of simulator might use concepts of general theories like the actor model, we prefer to use directly the VHDL-like instance as a starting point.

VHDL-like languages allow the programmer to define concurrent sections of code (processes), so that the order of these processes in the code is not important for the evaluation. A process described in the code is evaluated during the simulation every time an event occurs in one of its inputs (the sensitivity list). A process can be evaluated several times during the simulation. Inside each process, the programmer can use sequential code to have flexibility. An interesting feature is that data values can be defined as variables (that take a value immediately after assignment) or signals. Signals are special objects to hold data, but also events. In this way, when in a process a signal is assigned a new value, then the real assignment of this value is scheduled as an event for a future simulation iteration. In a simulation, the new value of signals is only visible after the end of the process. This is an interesting link with transactional memory.

We are looking for a first layer of software interface (of a higher level than a simple ISA) that fits the hardware characteristics of our proposal. Starting from the VHDL-like semantics and simulation, we propose the following modifications to support our first layer of programming interface:

1- Sequential view of execution in the concurrent part of VHDL-like programs: we want to preserve the intuitive vision of programmers of sequential code. The only condition we use is that signals propagate events in program order. The inputs activate a set of processes (hopefully concurrently), and then some events are produced on signals, but those events only activate processes downwards in the program order. Concurrency is obtained by the dependences indicated by signal names and the events on them. However the programmer's view is a sort of standard sequential program. If the programmer thinks parallel, then a high level of concurrency can be obtained. But, even with no care about parallelism, the model "discovers" the inherent parallelism (even in fully single threaded programs such as those of the SPEC benchmark there are abundant inherent parallelism). This feature is also being used in some dataflow-like proposals [20] [21].

2- In VHDL-like programs having two or more drivers for a single signal is not considered a good practice. Moreover, many dataflow models also have this restriction. This restriction may have sense for some cases, but it is too hard for others, leading to a lack of programming flexibility. There are many possibilities, and we should consider and evaluate them, but as a starting point we think to allow several drivers to the same signal name if the processes that act as drivers produce a final result as they were executed in a sequential way according to the program order. A simple example to clarify this is the following. Consider the simplistic code (in VHDL terminology) to find the maximum of an array of three values:

max<=a[0] **when** a[0] > max **else** max;

max<=a[1] **when** a[1] > max **else** max;

max<=a[2] **when** a[2] > max **else** max;

As we see, signal max has three drivers, but due to the propagation of signals in the direction of the program order (as described above in item 1), and the fact that max is also in the sensitivity list of the three processes (max is compared to the elements of the array), the three processes produce a final result as they were executed in sequential order. Therefore the processes are activated several times concurrently, but the net effect is a result as a sequential standard execution.

How do we connect this programming interface (VHDL-like event-driven) with our architecture proposal? We propose the following approach:

a) Code the application in a VHDL-like language (in the sense of event-driven with the modifications described above, but with all the needed high-level languages features for software productivity).

b) Run the program as a VHDL-like simulation: instead of compiling the program directly to an ISA, we feel that it might be more efficient to perform a "simulation" of the program, as it is done in VHDL simulation. This decouples the semantics of the high level language from the actual executing hardware. The simulator is in fact what in other context is called a run-time system (or in some cases an interpreter or virtual machine). But we prefer to preserve the notion of "simulation".

c) It can be argued that the "simulation" may lead to a highly inefficient execution (as it is usually the case in many interpreted languages not compiled in native ISA). But we think that this is the way to follow for abstracting the highly parallel hardware from the high level programming level. Therefore, we need a sort of "conventional" cores to "run" the simulation in parallel. This parallel engine for running the simulation could be customized for the efficient parallel implementation of an event-driven simulator. For this part we would use a conventional multithreaded model, and look for inspiration in existing proposals for parallel event-driven simulators.

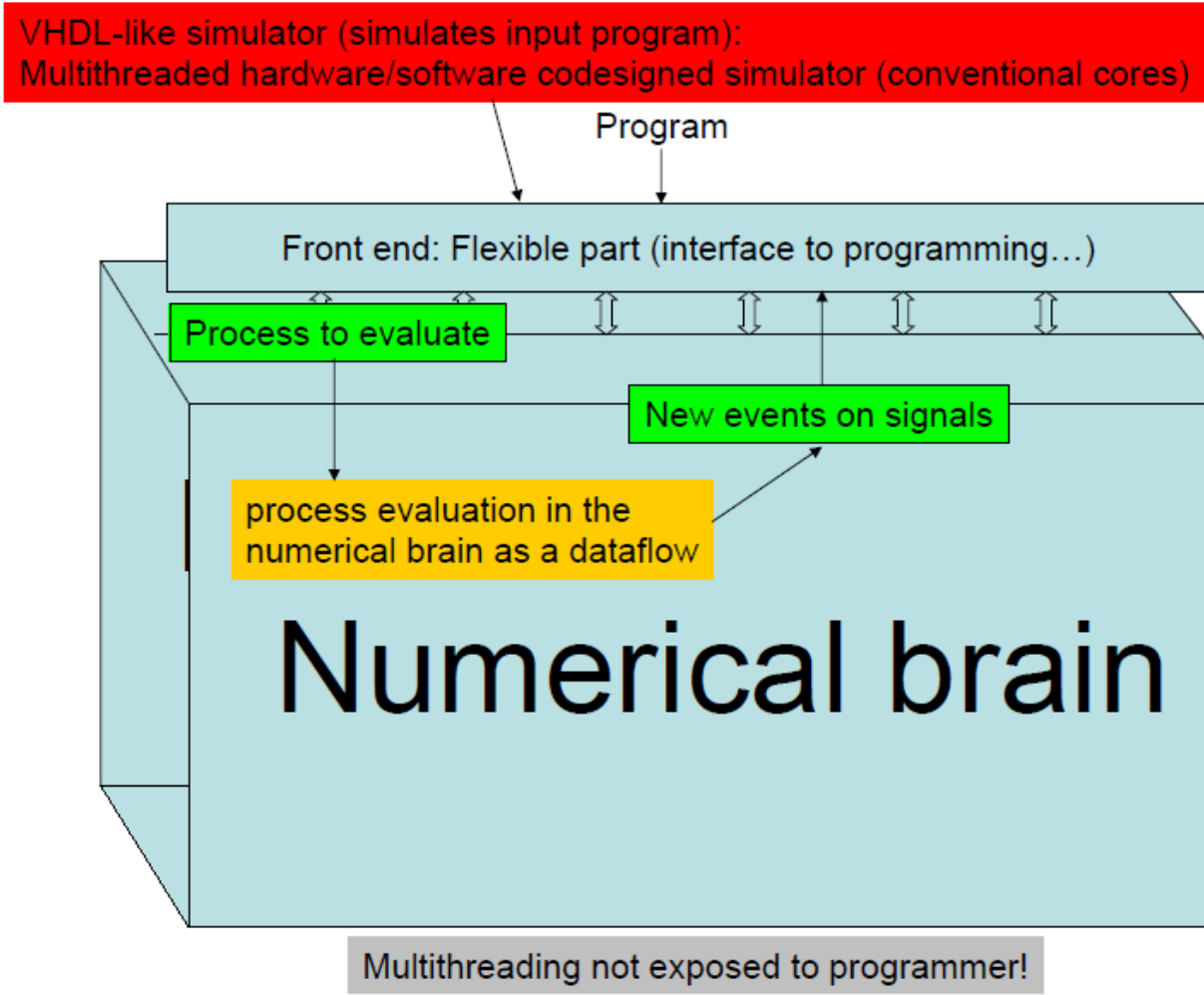

**Figure 3 : Overview of the ENBB microprocessor.**

d) The simulator uses the highly parallel packet switched network of arithmetic on-line resources described in previous sections, as a big coprocessor to perform the raw floating-point work of the application. This requires the conventional multithreaded cores to interact with the arithmetic network.

Therefore, our proposal is a heterogeneous system that is common in many proposals for future microprocessors. The novelty is that we propose to use the conventional cores to run only a specific program, the event driven simulator. Therefore this part is more a software/hardware co-design of a front-end for the arithmetic network. The interaction of the simulator (software and hardware) with the network is a key point that needs deep research. The basic unit of work will be the processes (in VHDL-like semantics), that are activated by signals during the simulation. The evaluation of those processes can be carried out in the arithmetic network, if the required work is hard enough in floating-point computations. In general, since the code inside a process is serial, a hardware dataflow would be established in the network (by means of control packets that reserve virtual channels). This is conceptually very similar to the work done by a VHDL hardware synthesis tool. The mapping of the dataflow in the network would be by means of the use of virtual channels in each switch, so that the input signals (of the sensitivity list) are injected (or moved) in the network to the appropriate switches to generate intermediate operands (variables in VHDL terminology that do not generate events), and produce output signals (as determined by the code). This is all done by chaining operations using the on-line arithmetic operators. Special control packets would make the reservation and configuration of virtual channels inside each switch, so that a virtual datapath is established in the network. Once the first digit of an output signal is produced, a control message indicating an event on a signal would be sent to the software/hardware "simulator", that would add the event to the pool of events for "future" activation of processes. Figure 3 illustrates the ENBB microprocessor composed by the parallel multithreaded event-driven simulator and the large numerical brain network. Note that most signals would be stored in the network, and they should be moved (possibly keeping a copy in its current place in some shared shadow flit buffers) to the appropriate network switch when the signal is required for the evaluation of a new process. As in the neocortex, we envision columns in several layers of the 3D network to set up a dataflow necessary for computing a process (neuromorphic architectures are also taking advantage of 3D integration). The column configuration may lead to better efficiency since the communication distances are reduced with respect to a pure 2D configuration. As in the neocortex, the "cells" of different columns will establish some communication channels due to signals needed that are stored elsewhere.

We expect to save a high amount of main memory requirements since variables inside the processes that are evaluated in the arithmetic network would not require explicit memory storage; these variables of the program just indicate a connection in the dataflow. Moreover, many signals would be stored in the network for further processing (activation of new processes). Processes may have parallel loops (very common in supercomputing simulations), so that the same dataflow "allocated" in the network can be used to process several or all the instances of the loop body.

There are several design decisions that should be taken for the efficient interaction between the arithmetic network and the software/hardware simulator front-end. Some of the main challenges are the following:

1) Most of the signals will be stored in the network. It is necessary to find the most effective way to control the movement of those signals to satisfy the computation needs: move the signal to a specific network location for dataflow evaluation of a new process, how to manage the situations when the same signal is needed in several dataflows, determine when a signal is no longer needed and it can be deleted, etc.

2) We would like not to have explicit main memory storage for signals that are stored in the network. However, it is necessary a scheme to have the flexibility of storing signals from the network to memory, to avoid the collapse of storage resources (virtual channels) in the network. If the size of the problem assigned to the microprocessor is appropriate for maximum concurrency (not oversized), and if it is easy to determine for most of the cases when a signal is no longer needed in the computation, the need for main memory storage for signals should be low.

3) We should incorporate effective program control flow (if-then-else type statements) in the dataflow execution model for the arithmetic network. One way of doing this is converting conditional execution into a full arithmetic expression (almost similar to what is done in predicated execution). This might be seem as a complex solution, but arithmetic is cheap in our case and allows a continuous flow in the execution.

4) We need to deal with loops when the number of iterations is data-dependent (both inside processes and outside processes). For dataflow execution we unroll the loops, but if the number of iterations is not known at compile time or at running time just before the loop, the simulator should implement same sort of dynamic unrolling if the body of the loop is worth enough for execution in the arithmetic network.

5) We need to prevent network saturation. The network should have moderate levels of utilization to avoid network saturation and excessive power dissipation. Having some inspiration from the neocortex would be of much interest. According to some evidence, the activation of neurons in the neocortex is relatively sparse. Probably we will need a very big network but with moderate utilization to satisfy both the performance requirements (if enough parallelism is found), and the power dissipation requirements (the "dark silicon" concept seems to be natural for power and energy efficient systems as the neocortex).

6) It will be a challenge to obtain scientific facts that demonstrate the effectiveness of the whole system.

7) The software/hardware simulator should be robust and enhance resilience, but it is necessary to address the specific engineering required for that.

As a final outcome of a research for this prototype it would be necessary to provide a scaling path for this microprocessor for successive technology nodes. We envision a path with a downscaling of frequencies and up scaling of total silicon area by means of a combination of up scaling of die areas and number of layers in the 3D chip. It would be also of interest to study the suitability of our architecture for future post-CMOS technologies.

*5- Methodology for Exploring the ENBB Design Space*

We identify naturally five big tasks to carry out a research for this microprocessor: i) design of the front-end hardware/software event-driven simulator and the semantics supported, ii) design of the big back-end arithmetic network coprocessor, iii) seamless integration of both parts, iv) run benchmarks and perform an evaluation of the system (energy efficiency, performance, reliability issues, etc.), with feedback for modifications of the design, v) demonstrate the hypothesis of "less main memory is more efficient" and vi) study a scaling path for future technologies.

We present an innovative microprocessor and its first layer of programming environment for a technology that will not exists until the end of the decade (or even technologies for the post ExaScale era). Moreover, implementing real hardware is a very time consuming process and very costly. Therefore, as it is usually done in many computer architecture projects, a research for the ENBB would rely on detailed simulation. An outcome of the research would be a detailed simulator of the microprocessor. A cycle accurate simulator would be needed that incorporates all the technological parameters so that scientific evidence can be found for the formulated hypothesis. In this scenario it is expected a high complexity in the simulation, so the simulator should be parallel and highly scalable (supercomputers should be used).

For the technology parameters, industry predictions should be used, and taking into account the possibility of different scenarios. Therefore, the simulations would be carried out for a set of different technological parameters (the processor should not be highly sensible to specific technological parameters). As it is standard in this kind of simulators, all the component modules should be parameterized in terms of energy, area, delay and reliability for a given technology parameters set. A key issue is the energy/delay/area of the interconnections. In principle, without an actual layout it is not possible to have accurate interconnection lengths. However, there exists models to estimate average interconnect length based on the fan-out and number of gates of the module. Moreover, certain reasonable geometric assumptions can be made, especially in regular structures, to estimate the layout interconnection length. Another issue would be to fully simulate a 3D structure, and the *memristor* like memories, since they are under development. It would be necessary a research of all the available prototypes to build a sound simulation model. It would be also necessary to incorporate simulation of (soft and hard) faults for deep reliability studies.

A bottom up methodology would be suitable for the research, by building basic modules and integrating them in a hierarchical way to have higher levels of complexity until getting the final full system simulator. For the design of the different parts, first, high level models for fast local exploration of design alternatives could be used, and then to develop the detailed models with the best candidate solutions. A final key issue would be to verify that the simulator performs correct simulations and that the models used to reproduce the behavior of actual hardware in the selected technology are accurate enough to provide scientific evidence.

*6- Conclusions*

Exascale systems will be a key element for simulations that are essential for economic growth in a highly technological innovation driven society. The current technology trends in microprocessors indicate that there might be a big energy wall in the near future. Different reports indicate the need of strong changes at all levels for Exascale computer systems. This position paper addresses this problem in the context of the microprocessor, the key element for performing floating-point work.

A conceptual microprocessor for floating-point computations was presented, useful for being a basic building block of Exascale systems. The proposed microprocessor architecture has a front end for programming interface based on the concept of event-driven simulation. The user program is executed as a event-driven simulation using a hardware/software co-designed simulator. This is the flexible part of the system. The back end exploits the concept of uniform topology: an interconnection network that incorporates seamlessly communication arithmetic and storage. Floating-point arithmetic is incorporated as on-line arithmetic operators in the output ports of the switches, and storage as virtual channels in the network. The front end carries out the "simulation" of the user program, and uses the arithmetic network for the hard floating-point work. The execution in the arithmetic network is of the dataflow type. We expect to reduce significantly the needs of main memory due to the execution model proposed, where conventional variables are just interconnections in the network or signals stored in the virtual channels. The memory needs required for the simulator could be integrated in the same chip as the microprocessor in a 3D structure. Moreover, we have the hypothesis that the problem size assigned to a microprocessor should allow maximum concurrency and it should not be oversized. This may lead to systems composed of microprocessors with main memory incorporated in 3D chips. We identified several challenges that a research for this microprocessor should address, and several hypothesis that should be demonstrated by means of scientific evidence.